\documentstyle[pra,aps,amssymb,amsmath,twocolumn]{revtex}

\def\vv{\underline}
\def\diag{\mathrm{diag}}
\renewcommand\vec[1]{{\mathbf #1}}
\def\mat[#1][#2]{\vspace{2ex}^{#1}\vspace{-2ex}\underline{\underline{#2}}}
\def\Gr{{\mathcal{G}}}

\begin{document}
\draft
\title{Properties of excitations in systems with a spinor Bose-Einstein
condensate}
\author{P\'eter Sz\'epfalusy$^{(1,2)}$ and Gergely Szirmai$^{(1)}$}
\address{$^{(1)}$Department of Physics of Complex Systems, Roland E\"otv\"os
University, P\'azm\'any P\'eter s\'et\'any 1/A, H-1117,
\\$^{(2)}$Research Institute for Solid State Physics and Optics of the
Hungarian Academy of Sciences, P.O.Box 49, H-1525}

\date{\today}
\maketitle

\begin{abstract}
General theory in case of homogenous Bose-Einstein condensed systems with
spinor condensate is presented for the correlation functions of density and
spin fluctuations and for the one-particle propagators as well. The random
phase approximation is investigated and the damping of the modes is
given in the intermediate temperature region. It is shown that the collective
and the one-particle excitation spectra do not coincide fully.
\end{abstract}


Since the realization of Bose-Einstein condensate in vapours of alkali-metal
atoms, there has been a tremendous development of the field concerning
experimental and theoretical aspects as well. Most of the works have considered
the gases in magnetic traps, where the spin degrees of freedom are frozen and
he particles behave as scalar ones. The confinement of the $^{23}Na$ gas in an
optical trap \cite{MIT} has opened a new avenue of investigations where the
spinor nature of the particle field plays a decisive role
\cite{Sea,ho,OM,LPB,HG1,HG2,HYip,HYin,CU}. The present paper is devoted also to
this problem considering a homogenous system.

Two states of systems with spinor condensates have been discovered
\cite{Sea,ho,OM}, namely the polar and the ferromagnetic ones and the
conditions of their appearance have been determined. It has also been shown
that the collective excitations possess different features in the two states
influenced also by a possible external magnetic field
\cite{ho,OM,HG2}. The Bogoliubov approximation,
which has been applied until now, does not take into account the interaction
between the particles in the condensate and the thermal cloud, leading to
undamped modes.

In this paper we start from the thermal Green's function formalism
\cite{FW,SG,Griffin}. First general relationships are derived for the density
and spin correlation functions by expressing them in terms of their proper
parts. Below $T_C$ the proper parts can be separated into singular and regular
contributions such that the diagrams contributing to former ones split by
cutting a single one-particle propagator. In lowest order the singular
diagrams alone lead to the Bogoliubov approximation for the modes. Treating the
regular part also in leading order damping of the modes and other qualitatively
new features emerge. The one-particle propagators
are also investigated and it is pointed out that the one-particle spectra
coincide with a part of the collective ones in the symmetry
breaking phase. Finally the damping of the modes are determined in the
intermediate temperature region.

The model used here is the usual one for particles with spin 1. The Hamiltonian
is
\begin{equation}\label{hamiltonian}
\begin{split}
H&=\sum_{\vec{k}}(e_{\vec{k}}-\mu)a^\dagger_r(\vec{k})a_r(\vec{k})+\\
&+\frac{1}{2}\mathop{\sum_{\vec{k}_1+\vec{k}_2=}}_{=\vec{k}_3+\vec{k}_4}
a^\dagger_{r'}(\vec{k}_1)a^\dagger_r(\vec{k}_2)V^{r's'}_{rs}a_s(\vec{k}_3)
a_{s'}(\vec{k}_4),
\end{split}
\end{equation}
where $e_{\vec{k}}=\hbar^2k^2/(2M)$ ($M$ is the mass of the atom), $\mu$ stands
for the chemical potential and the two particle interaction is given by
\begin{equation}\label{potential}
V^{r's'}_{rs}=c_n\delta_{rs}\delta_{r's'}+c_s(\vec{F})_{rs}(\vec{F})_{r's'}.
\end{equation}

Here the constants $c_n$ and $c_s$ are related to the s-wave scattering lengths
$a_0$ and $a_2$ in the total spin zero and two channel, respectively, by
$3c_n=4\pi\hbar^2a_n/M$ and $2c_s=4\pi\hbar^2a_s/M$, where $a_n=a_0+2a_2$ and
$a_s=2(a_2-a_0)/3$ \cite{Sea,ho,OM}. We assume that the system has unit volume.
$a_r(\vec{k})$ and $a^\dagger_r(\vec{k})$ are annihilation and creation
operators, respectively, for particles with momentum $\vec{k}$ and spin
component r. The representation $r=1,0,-1$ will be used, where the spin
operators are $F_z=\diag(1,0,-1)$, $(F_{\pm})_{rs}=(F_x)_{rs}\pm i(F_y)_{rs}=
\sqrt{2}(\delta_{\pm1,r}\delta_{0,s}+\delta_{0,r}\delta_{\pm1,s}$).
In Eq. \eqref{hamiltonian} and from now on summation over repeated
indices is understood.

To take into account the condensation the canonical
transformation
\begin{equation}\label{cantrf}
a_r(\vec{k})=b_r(\vec{k})+\delta_{k,0}\sqrt{N_0}\zeta_r
\end{equation}
is introduced, where $N_0$ denotes the number of particles in the zero momentum
state of spin projection r, forming the condensate and $\zeta$ is a normalized
spinor. $N_0$ is to be determined self-consistently.

As mentioned before there are two states according to the sign of the constant
$c_s$: The {\it polar state} ($c_s>0$), where $\left<\vec{F}\right>=0$,
$(\zeta^P)_r=\delta_{r,0}$, and the {\it ferromagnetic state} ($c_s<0$), where 
$\left<\vec{F}\right>^2=1$, $(\zeta^F)_r=\delta_{r,1}$.

We will deal with the following correlation functions
\begin{align}
D_{nn}(\vec{k}, \tau)=\Big< T_\tau\big[ n(\vec{k}, \tau)n(-\vec{k}, 0)
\big]\Big>,\label{dens}\\
D_{zz}(\vec{k}, \tau)=\Big< T_\tau\big[ {\mathcal{F}}_z(\vec{k}, \tau){
\mathcal{F}}_z(-\vec{k}, 0)\big]\Big>,\label{spindens}\\
D_{nz}(\vec{k}, \tau)=\Big< T_\tau\big[ n(\vec{k}, \tau){\mathcal
{F}}_z(-\vec{k}, 0)\big]\Big>,\label{spindens-dens}\\
D_{\pm\pm}(\vec{k},\tau)=\Big< T_\tau\big[ {\mathcal{F}}_\pm(\vec{k}, \tau)
{\mathcal{F}}_\mp(-\vec{k},0)\big]\Big>,\label{spinwdens}\\
D_{QQ}(\vec{k},\tau)=\Big< T_\tau\big[ {\mathcal{F}}_Q(\vec{k},\tau)
{\mathcal{F}}^\dagger_Q(-\vec{k},0)\big]\Big>.\label{quadspindens}
\end{align}
Here $n(\vec{k})=\sum_{\vec{q}}a^\dagger_r(\vec{k}+\vec{q})a_r(\vec{q})$
is the density operator, ${\mathcal{F}}_z(\vec{k})=\sum_{\vec{q}}
a^\dagger_r(\vec{k}+\vec{q})(F_z)_{rs}a_s(\vec{q})$, ${\mathcal{F}}_\pm(\vec{k})
=\sum_{\vec{q}}a^\dagger_r(\vec{k}+\vec{q})(F_\pm)_{rs}a_s(\vec{q})$,
${\mathcal{F}}_Q(\vec{k})=\sum_{\vec{q}}a^\dagger_r(\vec{k}+\vec{q})(F_+^2)
_{rs}a_s(\vec{q})$ are the appropriate spin-density operators. The
averaging is made over the grand canonical ensemble, $\tau$ is the imaginary
time and $T_{\tau}$ is the $\tau$ ordering operator \cite{FW}. 

We introduce a propagator from what the above ones can be easily
evaluated as
\begin{equation}\label{prop4}
D^{sr}_{r's'}(\vec{k},\tau)=\Big< T_\tau \big[ \sigma_{rs}(\vec{k},\tau)
\sigma_{s'r'}(-\vec{k},0)\big] \Big>,
\end{equation}
where $\sigma_{rs}(\vec{k})=\sum_{\vec{q}}a^\dagger_r(\vec{k}+\vec{q})
a_s(\vec{q})$.

The propagators are periodic in $\tau$ with period $\beta\hbar$, in
their Fourier-series the occurring frequencies are the Bose discrete Matsubara
frequencies $\omega_n=\frac{2\pi n}{\beta\hbar}$ \cite{FW}.
The propagators $D^{sr}_{r's'}(\vec{k}, i\omega_n)$ can be written in
terms of their proper parts $\Pi^{sr}_{r's'}(\vec{k},i\omega_n)$ (which do
not split by cutting a single interaction line):
\begin{equation}\label{geneq}
\begin{split}
D^{sr}_{r's'}(k,i\omega_n)=\Pi^{sr}_{r's'}(k,i\omega_n) +\\
+\Pi^{sr}_{ab}(k,i\omega_n)V^{ba}_{cd}D^{dc}_{r's'}(k,i\omega_n).
\end{split}
\end{equation}

Space reflection and time inversion symmetry yields $D^{sr}_{r's'}=
D^{r's'}_{sr}$. Spin conservation means that $r-s=r'-s'$. The same symmetry
relations hold for the proper parts. As a result Eq. \eqref{geneq} separates
into independent equations for specified spin transfers.

The first one (corresponding to zero spin transfer) reads as
$\mat[0][D]=\mat[0][\Pi]+\mat[0][\Pi]\cdot\mat[0][C]\cdot\mat[0][D]$,
where the following notation is introduced
\begin{equation}\label{d0mat}
\mat[0][D]=\left(
\renewcommand{\arraystretch}{1.2}
\begin{array}{c c c}
D^{++}_{++} & D^{++}_{00} & D^{++}_{--} \\
D^{00}_{++} & D^{00}_{00} & D^{00}_{--} \\
D^{--}_{++} & D^{--}_{00} & D^{--}_{--} \\
\end{array}
\right )
\end{equation}
and equivalently for $\mat[0][\Pi]$. Here and in the following $+,0,-$ is
used instead of $+1,0,-1$ for the spin components. The interaction matrix is
$\mat[0][C]=c_n\cdot \vv{\xi}_1\circ\vv{\xi}_1+c_s\cdot\vv{\xi}_2\circ
\vv{\xi}_2$, where $\vv{\xi}_1=(1,1,1)^T$ and $\vv{\xi}_2=
(1,0,-1)^T$. One can easily verify that after solving the matrix equation for
$\mat[0][D]$ the correlation functions defined in \eqref{dens}--\eqref{spindens-dens}
can be written in the following form: $D_{nn}=\vv{\xi}_1^T\cdot\mat[0][D]\cdot
\vv{\xi}_1$, $D_{zz}=\vv{\xi}_2^T\cdot\mat[0][D]\cdot\vv{\xi}_2$ and $D_{nz}=
\vv{\xi}_1^T\cdot\mat[0][D]\cdot\vv{\xi}_2$. Since $\mat[0][D]$ is a symmetric
matrix (due to the above discussed symmetry relations) $D_{nz}=D_{zn}$.

The second equation (corresponding to plus one spin transfer) is
$\mat[+][D]=\mat[+][\Pi]+\mat[+][\Pi]\cdot\mat[+][C]\cdot\mat[+][D]$,
where
\begin{equation}\label{d+mat}
\mat[+][D]=\left(
\renewcommand{\arraystretch}{1.2}
\begin{array}{c c}
D^{0+}_{+0} & D^{0+}_{0-}\\
D^{-0}_{+0} & D^{-0}_{0-}\\
\end{array}
\right)
\end{equation}
and its proper part matrix $\mat[+][\Pi]$ is similarly defined. The
interaction here can be written in the form $\mat[+][C]=c_s\cdot\vv{\chi}
\circ\vv{\chi}$, where $\vv{\chi}=(1,1)^T$. Furthermore
$D_{++}=\vv{\chi}^T\cdot\mat[+][D]\cdot\vv{\chi}$.

The equation for the negative spin transfer (for $\mat[-][D]$) can be
calculated in the same way. The equation corresponding to the plus two spin
transfer is trivial, since the interaction matrix becomes zero, consequently
this propagator is already proper: $D^{-+}_{+-}=\Pi^{-+}_{+-}$.

The excitation energies and their widths can be found by analytic
continuation on the omega plane in the usual way \cite{FW} and are determined
by the condition
$\det(\mat[][I]-\mat[\genfrac{}{}{0pt}{}{0}{+}][\Pi]\cdot\mat[
\genfrac{}{}{0pt}{}{0}{+}][C])=0$,
where $\mat[][I]$ is the unit matrix of rank three and two in the
case of zero and plus one spin transfer, respectively.
The simplest possible approximation is to keep only the leading singular
contributions for the proper parts. Such an approach is equivalent to the
Bogoliubov approximation. The next simplest calculation is to keep also the
leading regular diagrams as well, giving the contribution $\Pi_0\cdot
\mat[][I]$. Here $\Pi_0$ is the contribution of the bubble diagram
\begin{equation}\label{bubble}
\Pi_0(\vec{k},i\omega_n)=\int\frac{d^3q}{(2\pi)^3}
\frac{n^0(\vec{k}+\vec{q})-n^0(\vec{k})}{\hbar i\omega_n-(e_{\vec{k}+\vec{q}}-
e_{\vec{k}})},
\end{equation}
where $n^0(\vec{k})$ is the Bose-Einstein distribution function for particles
of energy $e_{\vec{k}}$.

The calculation in this random phase approximation can be carried out
separately for the two states.

In the {\it polar state} the leading singular contributions are
\begin{subequations}\label{bogoeq}
\begin{align}
&\mat[0][\Pi]_B=\Pi_S\cdot\vv{\zeta}\circ\vv{\zeta},\label{bogodens}\\
&\mat[+][\Pi]^P_B=\left(
\begin{array}{c c}
\Pi_-^P & 0\\
0 & \Pi_+^P
\end{array}
\right).\label{bogospol}
\end{align}
\end{subequations}

In Eq. \eqref{bogoeq} $\Pi_S=2N_0\hbar^{-2}e_{\vec{k}}/\big[ (i\omega_n)^2
-\hbar^{-2}e_{\vec{k}}^2 \big]$, $\Pi_{\pm}= N_0\hbar^{-1}/(\pm i\omega_n-
\hbar^{-1}e_{\vec{k}})$. 

Using $\mat[0][\Pi]=\Pi_0\cdot\mat[0][I]+\mat[0][\Pi]_B$ the determinant
$\det(\mat[][I]-\mat[0][\Pi]\cdot\mat[0][C])=(1-2c_s\Pi_0)(1-3c_n\Pi_0-c_n
\Pi_S)$. These spectra separate into two parts, corresponding to the density
and the spin density mode described by
\begin{equation}\label{bubdenspol}
D_{nn}=\frac{3\Pi_0+\Pi_S}{1-3c_n\Pi_0-c_n\Pi_S}
\end{equation}
and
\begin{equation}\label{bubspindenspol}
D_{zz}=\frac{2\Pi_0}{1-2c_s\Pi_0},
\end{equation}
respectively.

The density and the spin fluctuations are not coupled in this state
($D_{nz}=0$) and the spin fluctuations are created only by the thermal cloud.

For the spin wave mode $\det(\mat[][I]-\mat[+][\Pi]\cdot\mat[+][C])=1-2c_s\Pi_0
-c_s\Pi_S$ and $D_{++}=\vv{\chi}^T\cdot\mat[+][D]\cdot\vv{\chi}$ results in
\begin{equation}\label{+mode}
D_{++}=\frac{2\Pi_0+\Pi_S}{1-2c_s\Pi_0-c_s\Pi_S}.
\end{equation} 

The same result holds for the $D_{--}$ spin--wave mode. The quadrupolar
spin--density fluctuation propagator is simple since $D_{QQ}=4D^{-+}_{+-}= 4
\Pi_0$, which means that this belongs also only to the non--condensate.
Keeping only the singular contributions $\mat[][\Pi]_B$ the poles of $D_{nn}$
and $D_{++}$ are at frequencies $\omega^P_{nn}=\hbar^{-1}\sqrt{e_\vec{k}
(e_\vec{k}+2N_0c_n)}$, $\omega^P_{++}=\hbar^{-1}\sqrt{e_\vec{k}(e_\vec{k}+
2N_0c_s)}$, respectively, in agreement with the modes known from previous
calculations\cite{ho,OM}.

We do not go into any detail concerning the {\it ferromagnetic state} and
give the result only for the density fluctuations.
The determinant $\Delta\equiv \det(\mat[][I]-\mat[0][\Pi]\cdot\mat[0][C])=(1-3c_n\Pi_0-c_n
\Pi_S)(1-2c_s\Pi_0-c_s\Pi_S)-c_nc_s\Pi_S^2$ does not factorize into two parts,
which means that the density and the spin density autocorrelation functions
(defined in Eqs. \eqref{dens} and \eqref{spindens}) are coupled, so the cross
correlation function (defined in Eq. \eqref{spindens-dens}) is not zero
either. One can find
\begin{subequations}\label{0modesfer}
\begin{align}
&D_{nn}=\frac{3\Pi_0(1-2c_s\Pi_0)+\Pi_S(1-5c_s\Pi_0)}{\Delta},
\label{bubsdensfer}\\
&D_{zz}=\frac{2\Pi_0(1-3c_n\Pi_0)+\Pi_S(1-5c_n\Pi_0)}{\Delta},
\label{bubspindensfer}\\
&D_{nz}=D_{zn}=\frac{\Pi_S}{\Delta},\label{bubdensspindensfer}
\end{align}
\end{subequations}
where $\Delta$ is the above given determinant.

We note that the results of the Bogoliubov approximation can be simply
regained here as well by setting $\Pi_0$ to zero, leading to the eigenfrequency
$\omega^F_n=\hbar^{-1}\sqrt{e_\vec{k}[e_\vec{k}+2N_0(c_n+c_s)]}$ in agreement
with\cite{ho}, \cite{OM}.

Before further discussing the spectra of the correlation functions
\eqref{dens}--\eqref{quadspindens} we present the structure of the thermal
Green's functions defined by
\begin{equation}\label{greendef}
\Gr^{rs}_{\alpha\beta}(\vec{k},\tau)=-\Big< T_{\tau}\big[b^{\beta}_r(\vec{k},
\tau)b^{\alpha \dagger}_s(\vec{k},0)\big] \Big>.
\end{equation}
The Roman indices are standing for the spin component, while the Greek ones
(taken as +1, -1) are for distinguishing between the normal and anomalous
Green's functions \cite{FW,SG,Griffin,SzK}, using the notation $b^1_r(\vec{k})=b_r(\vec{k})$,
$b^{-1}_r(\vec{k})=b^\dagger_r(-\vec{k})$.

The generalized Dyson-Beliaev equation for the Green's function reads as
$\Gr^{rs}_{\alpha\beta}=\Gr^{rs}_{_{(0)}\alpha\beta}+\Gr^{rs'}_{_{(0)}\alpha
\gamma}\Sigma^{s's''}_{\gamma\delta}\Gr^{s''s}_{\delta\beta}$,
where $\Sigma^{rs}_{\alpha\beta}$ is the self energy, the sum of those diagrams
with two external points which can not be split by cutting a single particle
line. The free propagator in our case is $\Gr^{rs}_{_{(0)}\alpha\beta}(\vec{k},
i\omega_n)=\delta_{rs}\delta_{\alpha\beta}/(\alpha i \omega_n-\hbar^{-1}
e_{\vec{k}})$.

Time reversal and space reflection invariance mean that
$\Gr^{rs}_{\alpha\beta}(\vec{k}, i\omega_n)=\Gr^{sr}_{\beta\alpha}(\vec{k},
i\omega_n)=\Gr^{rs} _{-\alpha,-\beta}(\vec{k}, -i\omega_n)$. Spin preservation
requires for $\Gr^{rs}_{\alpha\beta}$ that $\alpha r-\beta s=(\alpha-\beta)
\sigma$, where $\sigma$ is the spin of a condensate particle. This
structure can be symbolized for the polar state by
\begin{displaymath}
\Gr^{rs}_{11}:
\begin{array}{|c|c|c|}
\hline
\blacksquare & 0 & 0\\
\hline
0 & \blacksquare & 0\\
\hline
0 & 0 & \blacksquare\\
\hline
\end{array}\quad, \quad
\Gr^{rs}_{-11}:
\begin{array}{|c|c|c|}
\hline
0& 0 & \blacksquare\\
\hline
0 & \blacksquare & 0\\
\hline
\blacksquare & 0 & 0\\
\hline
\end{array}\quad,
\end{displaymath}
and for the ferromagnetic state by
\begin{displaymath}
\Gr^{rs}_{11}:
\begin{array}{|c|c|c|}
\hline
\blacksquare & 0 & 0\\
\hline
0 & \blacksquare & 0\\
\hline
0 & 0 & \blacksquare\\
\hline
\end{array}\quad, \quad
\Gr^{rs}_{-11}:
\begin{array}{|c|@{\hspace{1.65mm}}c@{\hspace{1.65mm}}|@{\hspace{1.65mm}}c|}
\hline
\blacksquare& 0 & 0\\
\hline
0 & 0 & 0\\
\hline
0 & 0 & 0\\
\hline
\end{array}\quad,
\end{displaymath}
where a black box refers to a nonzero element.
The same structure holds for the self energies as well.  

The dielectric formalism worked out for scalar Bose particles
\cite{Griffin,SzK,MW,KSz,PG,WG,BSz} can be generalized for the present case. In
this framework the expressions for the correlation functions $D_{nn}$, $D_{++}$
in the polar state and $D_{nn}$ ($D_{zz}$, $D_{nz}$) in the ferromagnetic state
can be rearranged so that they have the same denominator as the propagators
$\Gr^{00}_{11}$ ($\Gr^{00}_{1-1}$), $\Gr^{11}_{11}$ ($\Gr^{1-1}_{1-1}$) in the
polar state and $\Gr^{11}_{11}$ ($\Gr^{11}_{1-1}$)  in the ferromagnetic state,
respectively. (We listed only those relevant to our considerations here.)
Consequently the above pairs share the same excitation spectra.

We are not going to present the general theory here, instead we demonstrate
this property in the RPA. The interaction potential defined
in Eq. \eqref{potential} is renormalized to
\begin{equation}\label{renpotential}
\begin{split}
W^{r's'}_{rs}(\vec{k},i\omega_n)=\frac{c_n}{1-3c_n\Pi_0(\vec{k},i\omega_n)}
\delta_{rs}\delta_{r's'}+\\
+\frac{c_s}{1-2c_s\Pi_0(\vec{k},i\omega_n)}(\vec{F})_{rs}(\vec{F})_{r's'}.
\end{split}
\end{equation}
The diagrams contributing to the self-energies are drawn in Fig \ref{selfen}.

For the polar state the resulting Green's functions are
\begin{equation}\label{bubgrpol}
\Gr^{{00}\atop{++}}_{11}=\frac{(1-c\Pi)(i\omega_n+\hbar^{-1}e_{\vec{k}})+
\hbar^{-1}N_0c}{[(i\omega_n)^2-\hbar^{-2}e_{\vec{k}}^2](1-c\Pi)-2\hbar^{-1}
e_{\vec{k}}N_0c},
\end{equation}
where $c=c_n$, $\Pi=3\Pi_0$ and $c=c_s$, $\Pi=2\Pi_0$ for $\Gr^{00}_{11}$ and
$\Gr^{++}_{11}$, respectively.

For the discussed mode in the ferromagnetic state the Green's function is
\begin{equation}\label{bubgrfer}
\Gr^{++}_{11}=\frac{\hbar(i\hbar\omega_n+e_{\vec{k}})(1-3c_n\Pi_0)(1-2c_s\Pi_0)
+\hbar N_0\rho}{[(i\hbar\omega_n)^2+e_\vec{k}^2](1-3c_n\Pi_0)(1-2c_s\Pi_0)
+2N_0e_{\vec{k}}\rho}.
\end{equation}
Here the abbreviation $\rho=c_n+c_s-5c_nc_s\Pi_0$ is introduced to simplify
the equation.

Comparison of Eq. \eqref{bubgrpol} with \eqref{bubdenspol} and \eqref{+mode}
and furthermore \eqref{bubgrfer} with \eqref{0modesfer} illustrate the statement
above. Note that $D^P_{zz}$ and $D^P_{QQ}$ do not have counterparts among the
Green's functions, which means that the collective and one-particle spectra do
not coincide completely.

Now we turn our attention evaluating the spectra. It is useful to introduce
characteristic lengths, namely the thermal wavelength $\lambda=\hbar/(2Mk_BT)^{1/2}
$, the correlation length $\xi'=Mk_BT/(4\pi\hbar^2N_0)$ governing the critical
fluctuations and the mean field (Bogoliubov) coherence lengths $\xi^B_{n,s}=
\hbar/(4MN_0c_{n,s})^{1/2}$.

We consider the intermediate temperature region,
where $\xi^B\gg \lambda,\xi'$ and the contribution of the bubble graph can
be treated as a perturbation. To leading order shift in the energy can be
neglected and the main effect is the damping of the eigenmodes due to their
interaction with the thermal cloud represented by the bubble graph.
This is a Landau-type damping which is the dominating process at such
temperatures. $\Pi_0$ has been analyzed in detail in \cite{SzK}. In the
temperature region considered and for linear dispersion
$\Pi_0$ can be approximated by $i[M/(4\pi\hbar^2k\lambda^2)]\cdot
\ln[(\hbar\omega-e_{\vec{k}})/(\hbar\omega+e_{\vec{k}})]$.
The spectra can be written in the form $\omega-i\gamma$, where
$\omega$ is the frequency in the Bogoliubov approximation for small k limit
and $\gamma$ represents the damping if $\omega$ is not zero.
In the {\it polar state} ($\gamma^P_{nn}/\omega^P_{nn})=(3\xi'/2\xi^B_n)$
and ($\gamma^P_{++}/\omega^P_{++})=(2\xi'/2\xi^B_s)$. Furthermore $D_{zz}$ has a
pole at $\omega^P_{zz}=0$, $\gamma^P_{zz}=2\gamma^P_{++}$.
In the {\it ferromagnetic state} $\gamma^F_{nn}=\gamma^F_{zz}=\gamma^F_{nz}$
and $\gamma^F_{nn}/\omega^F_{nn}=\xi'[3(\xi^B_s)^2+2(\xi^B_n)^2]/[2\xi^B_n
\xi^B_s\sqrt{(\xi^B_s)^2+(\xi^B_n)^2}]$.

In the experiments made with sodium atoms \cite{MIT}, the scattering lengths
are $a_0=(46\pm5)a_B$, $a_2=(52\pm5)a_B$ (see e.g. \cite{ho}), where $a_B$ is
the Bohr radius, which means that $c_n>0$ i.e. the polar state is realized in
equilibrium and in the absence of an external magnetic field. For an estimate
the temperature can be taken to be of the order of $10^{-7}K$ and the
condensate density is about of $3\cdot10^{-20}m^{-3}$. With these parameters
$\gamma^P_n/\omega^P_n\approx1,2\cdot10^{-2}$ and
$\gamma^P_+/\omega^P_+\approx1,6\cdot10^{-3}$, showing that the spin wave is
much less damped than the density fluctuations.

Finally some remarks are appropriate about the relevance of calculations for
homogenous system to the Bose gas in a trap, beyond the obvious usefulness
namely that they can serve as a guide. As far as the damping of the modes is
concerned one can refer to the fact that the simple expression obtained in RPA
\cite{SG,SzK} has proved to be quantitatively a good approximation
\cite{BSz,liu,PS,FSW} in case of gases in magnetic traps, even when compared to
the experimental findings \cite{Jea}. One may expect a similar situation here.
Moreover, the results obtained for the homogenous system are relevant in the
local density approximation for the gas in a trap. Namely by exciting density
perturbations much smaller than the size of the condensate the local speed of
the quasiparticle (zeroth sound) has been measured and its density dependence
has been found in agreement with the bulk result \cite{Aea}. One can hope that
such experiments can be extended to spin waves.

The present work has been partially supported by the Hungarian Academy of
Sciences under Grant No. AKP 9820 2,2 and by the Hungarian Ministry of Education
under Grant No. FKFP 0159/1997.

\newpage
\begin{figure}[h]
\caption{Self energy contributions in RPA, the solid line denotes a
one-particle free propagator, a double--dashed--line denotes the renormalized
interaction and the circle is representing a conensate particle.
}\label{selfen}
\end{figure}

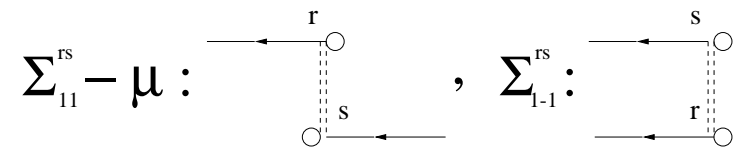
\end{document}